\documentclass[preprint]{acmart}

\AtBeginDocument{%
  \providecommand\BibTeX{{%
    \normalfont B\kern-0.5em{\scshape i\kern-0.25em b}\kern-0.8em\TeX}}}

\setcopyright{acmcopyright}
\copyrightyear{2021}
\acmYear{2022}

\acmConference[CHI'22]{CHI 2022: ACM CHI Conference on Human Factors in Computing Systems}{April 30 -- May 6, 2022}{New Orleans, LA}
\acmBooktitle{CHI 2022: ACM CHI Conference on Human Factors in Computing Systems, April 30 -- May 6, 2022, New Orleans, LA}

\begin{document}

\title[Uplifting Interviews in Social Science with Individual Data Visualization]{Uplifting Interviews in Social Science with Individual Data Visualization: the case of Music Listening}
\author{Robin Cura}
\authornote{Corresponding author}
\email{robin.cura@parisgeo.cnrs.fr}
\orcid{0000-0001-5926-1828}
\affiliation{%
  \institution{CNRS, UMR Géographie-cités, Paris}
  \country{France}
}

\author{Amélie Beaumont}
\affiliation{%
  \institution{Université Paris 1 Panthéon-Sorbonne, UMR Cresppa-CSU, Paris}
  \country{France}
}

\author{Jean-Samuel Beuscart}
\author{Samuel Coavoux}
\orcid{0000-0001-7991-3555}
\affiliation{%
  \institution{Orange Labs, Paris}
  \country{France}
}

\author{Noé Latreille de Fozières}
\affiliation{%
  \institution{CNRS, UMR Géographie-cités, Paris}
  \country{France}
}

\author{Brenda Le Bigot}
\affiliation{%
  \institution{Université de Poitiers, UMR MIGRINTER, Poitiers}
  \country{France}
}

\author{Yann Renisio}
\orcid{0000-0001-9415-6193}
\affiliation{%
  \institution{CNRS, UMR Observatoire Sociologique du Changement, Paris}
  \country{France}
}

\author{Manuel Moussallam}
\orcid{0000-0003-0886-5423}
\affiliation{%
  \institution{Deezer Research}
  \country{France}}
\email{mmoussallam@deezer.com}

\author{Thomas Louail}
\orcid{0000-0001-8563-6881}
\email{thomas.louail@cnrs.fr}
\affiliation{%
  \institution{CNRS, UMR Géographie-cités, Paris}
  \country{France}
}
\renewcommand{\shortauthors}{Cura et al.}

\begin{abstract}
Collecting accurate and fine-grain information about the music people like, dislike and actually listen to has long been a challenge for sociologists. As millions of people now use online music streaming services, research can build upon the individual listening history data that are collected by these platforms. Individual interviews in particular can benefit from such data, by allowing the interviewers to immerse themselves in the musical universe of consenting respondents, and thus ask them contextualized questions and get more precise answers. Designing a visual exploration tool allowing such an immersion is however difficult, because of the volume and heterogeneity of the listening data, the unequal "visual literacy" of the prospective users, or the interviewers' potential lack of knowledge of the music listened to by the respondents. In this case study we discuss the design and evaluation of such a tool. Designed with social scientists, its purpose is to help them in preparing and conducting semi-structured interviews that address various aspects of the listening experience. It was evaluated during thirty interviews with consenting users of a streaming platform in France. 
\end{abstract}



\begin{teaserfigure}
    \includegraphics[width=\textwidth]{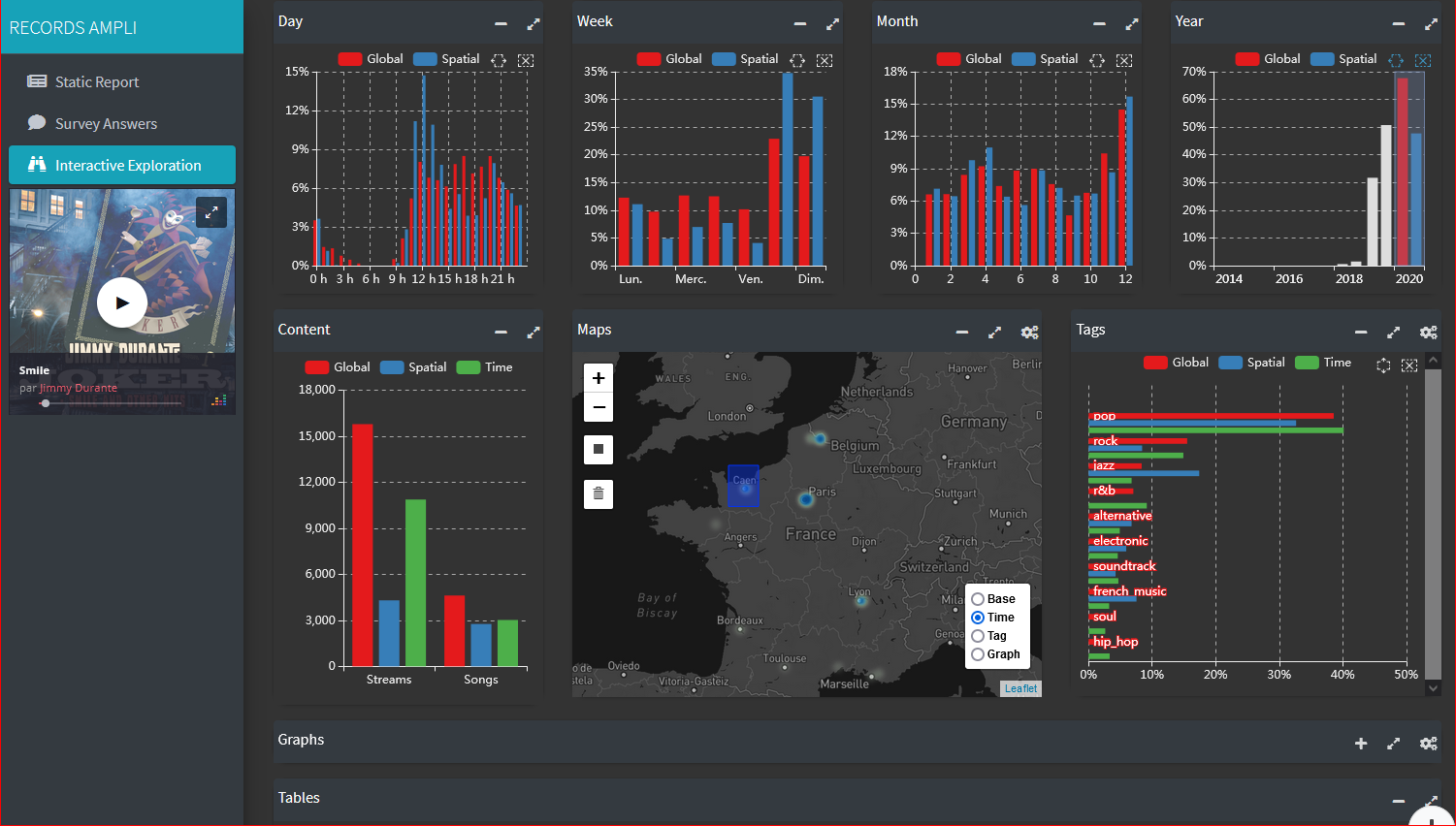}
    \caption{An example of AMPLI's exploration part. Here, there is both a graphical selection brushed in France Normandy region, and selecting only the year 2020 of the data. We can see that the streams originating from this area comprised more Jazz music than usual for this interviewee (as illustrated by the "Tags" bar-plot in the bottom-right). The song the interviewee played the most in this area and during that time-frame was "Smile" by Jimmy Durante (as seen on the music player widget on the left menu).}
    \label{fig:ampli_exploration}
    \Description{This image shows the exploratory part of AMPLI graphical interface. It displays a song being played (Jimmy Durante's "Smile"), and various interactive bar-plots showing the temporal and spatial dimensions of a given Deezer user's streams. Both a temporal selection (streams from 2020 only) and a spatial selection (only streams located in Normandy) are made, and the visual elements of the window reflect the differences in musical genres listened to by the interviewee when these filters apply.}
    \clearpage
\end{teaserfigure}

\maketitle

\section{Introduction}
This case study is part of an interdisciplinary project (www.records.huma-num.fr) in which social scientists and computer scientists, in collaboration with the research department of a music streaming platform, are investigating the social stratification of music listening practices in France. As part of this research, individual interviews are conducted with users of the platform. These are sampled ($n \approx 100$) among the respondents of a questionnaire survey ($n \approx 15000$) who gave their written consent (i) to participate in a recorded interviewed and (ii) for their several-years listening history data on the platform to be linked with their answers to the questionnaire survey. The interviews browse a number of subjects related to music listening and music taste. These include the interviewee's actual use of the streaming service ; the spatial and temporal contexts in which he/she listens to music ; his/her music taste and distaste, preferences for certain genres and artists, etc. ; and finally his/her use and appreciation of the platform's recommender system. To assist the social scientists in preparing their interviews -- that is to develop some first insights of the streaming practices and the music taste of the persons they will interview -- we wished to design an exploratory data analytics tool. This tool, named AMPLI -- for Amplifying Music listening Practices studies with Logs-augmented Interviews -- should allow interviewers to visualize the answers of the respondents to the survey, and it should also allow to query and visualize the several-year listening history data of the respondents collected by the streaming platform. These data contain many bits of information for each single time-stamped stream, including the song/artist/album listened to, if the song was liked/banned/skipped or not, the estimated location of the user, the type of device used (was it a mobile or a tablet or a web browser?), whether the stream was 'organic' (e.g. playlist, search) or recommended (algorithmic recommendation or editorially edited platform playlist), etc. We intended the primary users of this tool to be the social scientists conducting the qualitative interviews, and in the following we will refer to them as \emph{the/our users}. 

In the next section, we give a general overview of the relevant research literature. We then present our users' needs and the constraints we identified. Then we present the design and interactivity choices we made to provide the users with a tool suited for the anticipated use cases. Finally, we discuss the feedback we received from the users after a first round of thirty individual interviews, focusing on the unexpected uses that were made of AMPLI.

\section{A synthetic review of music data visualization\label{sec:review}}

The design of AMPLI took place at the crossroads of different research fields, including Computer-Human Interaction and Data visualization. Our goal was to build a tool that would allow to visualize and to interact with musical data. From the typology introduced in \citet{khulusi_survey_2020}'s recent literature review, we identified two particular tasks of the "visualization of musical collections" category : the "Explorative analysis" and the "analysis of listening statistics" tasks.
For both tasks, Dominikus Baur and his team made several proposals~\cite{baur_pulling_2009,baur_songs_2011,baur_arcsfm_2012}.
In particular, \citet{baur_arcsfm_2012} designed \textit{arcs.fm}, an application to visually analyze Last.fm users' listening history data. Its purpose is to support a conversation between two Last.fm users who can visually compare their listening activity on the service. They can thus easily identify the music taste and distaste they have in common, and talk about it. While several visual proposals in \textit{arcs.fm} are undoubtedly inspirational, its use requires at least a basic knowledge of the songs and artists that were listened to. This makes it unsuitable for our own use case because our users visualize someone else's listening history data, not their own. Indeed, one of the main requirements of the social scientists was to gain insights and develop a general intuition of what the interviewees listen to, that may be very different from what themselves listen to.

The existing literature on how to visualize music streaming data can be roughly divided in two parts, on one side visualizations and interaction schemes that are specific to musical data, and on the other "generic" visualizations and interactions. While the former~\cite{torrens_visualizing_2004,zhang_visual_2017,ono_similarity_2015} results in visual synthesis of music data that are more expressive, we had to consider the unequal "visual literacy" of our users who may be not used to interacting with state-of-the-art visualization methods. Consequently, we decided to rely on simpler, standard data visualization components such as histograms, bar-plots, line-plots, geographic maps and interactive graphs. Each of these plots alone carries less information, but their assembly and cross-interactivity allow for complex visual queries and information retrieval. This approach was followed in~\cite{baur_arcsfm_2012,chen_gaining_2010,dias_interactive_2012} and also by \citet{wirfs-brock_giving_2020}~in an experiment where a few interviewees were exposed to "silent data" extracted from their Spotify listening history in order to help them defining how they would command a voice assistant to choose music. Although our users are different -- namely social scientists conducting interviews -- this study is inspirational in using simple plots that are easy to understand for any user, even the less "visually literate".

Another body of work that proved useful relates to the use of personal data and visualization interfaces during the interview process. Up to our knowledge social science research on this issue has been limited, although the increasing availability of individual digital traces should soon support new methodological proposals, and hopefully the emergence of standards and good practices.
\citet{dubois_trace_2015}'s "Trace Interviews" may be one of the very first attempts to frame the various problems encountered when using data visualization as a companion tool for social research interviews. The authors collected publicly available digital footprints that were produced by political representatives as part of their activity on the internet (including tweets and wikipedia edits), and then invited these persons for an interview. During the interview, people were presented with different visualizations of their data and were asked to discuss both the content and the form. As the authors phrase it:
\begin{quote}
    "Trace interviewing provides opportunities to (1) enhance recall and generate an understanding of an actor’s decision-making process, (2) validate trace data-generated results, (3) address the data joining problem as interviewees explain traces on multiple platforms, and (4) respond to ethical problems relating to the use of personal data without consent or involvement by users."
\end{quote}
A major difference between this experiment and ours is that the interviewees of \cite{dubois_trace_2015} were selected according to some content they deliberately wrote on the internet. More precisely, they were selected because what they wrote was interesting enough, both in quantity and quality, to identify topics that could be discussed and enriched through a qualitative interview. Furthermore, internet data collected in this study -- and the associated visualizations -- were mono-dimensional: the two case studies mentioned in this paper rely upon a couple of static plots, illustrating a well-defined issue that does not require interactivity or data drill-down to be discussed. In our case the constraints are different, and to the opposite our goal was to assemble dozens of plots to embrace the multiple aspects of the experience of navigating and listening to music on a streaming platform -- as illustrated by Figure~\ref{fig:ampli_exploration} in the introduction. 

Our case study bears more similarity with a recent study \cite{moore_interview_2021,moore_exploring_2021} where personal data availability was not the initial trigger but came as a supplementary material for enriching otherwise scheduled interviews. 
Here the authors developed a "middle ground approach" of personal data visualization, lying between what they call "lightweights methods -- designing sketches or building data physicalization~\cite{jansen_opportunities_2015} of a user personal data -- and "heavyweight methods" -- where tools are designed to let users answer specific and predefined questions about their own data. By showing interviewees the computational notebook they used to analyze their data, the authors collect insights on how interviewees engage with their own logged behavior when visualizing the resulting data.

\section{Designing an application for "augmented" interviews}

To design AMPLI we followed three steps identified by ~\citet{oppermann_data-first_2020} in their "data-first design studies" proposal. The first step consisted in demonstrating to our intended users -- social scientists who were to conduct the interviews -- the potential of listening history data to retrieve information that would prove useful to prepare the interviews, and that would get the interviewees to clarify some points that otherwise would likely not be raised. We did so by showing them many examples of possible visualizations. This is the \textit{elicitation} phase. In the second step, we gathered their needs, constraints and ideas in order to quickly elaborate a first prototype of the tool. This is the \textit{winnowing} phase. The third step consisted first in observing the actual uses of the software during interviews -- different users made a different use of it --, and then in focusing on how to gather relevant feedback that would answer the following questions: what was missing, what features could be added, and maybe more importantly how exactly the application was used in "real-life" conditions -- that is during the interviews. This is the \textit{deployment} phase.

\begin{figure*}[!htb]
    \centering
    \includegraphics[width=.6\textwidth]{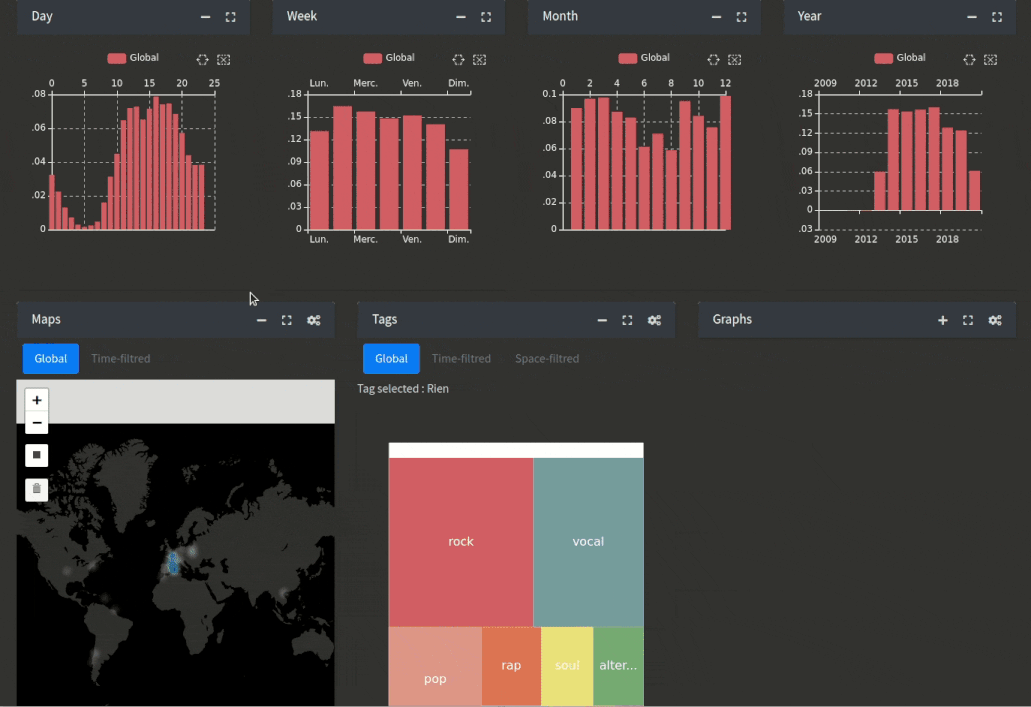}
    \caption{\label{fig:proto_ampli}Early prototype of AMPLI. Its purpose was to convince the prospective users -- social scientists who conduct individual interviews -- that the visual exploration of people's listening history data had great potential to suggest hot topics for the forthcoming interview, and to make people talk about their music listening practice, provide details and nuances, etc. and finally improve the quality of the information collected during the interview.}
    \Description{This figure displays an early prototype of the AMPLI application. It shows the streams of a Deezer user, and using bar-plots, their temporal occurrence in terms of hourly, daily, monthly and yearly distribution. On the second row of the application, this interface displays a geographical heat-map of where the streams were made, and a tree-map of the musical genres that those streams corresponds to.}
\end{figure*}

\begin{itemize}
    \item \textbf{Elicitation} : we developed a visual analytics prototype to make our users aware of the diversity and amount of information that are associated to each individual stream. This prototype allowed them to explore the several-year listening history data from a single random user of the streaming platform (see figure~\ref{fig:proto_ampli}). It focused on the spatial and temporal dimensions of the streaming data, and on the music genres that were listened to. The main purpose was to illustrate how interactive cross-filtering of the space-time variables -- approximate geographical location retrieved from the IP address of time-stamped streams -- could help in uncovering interesting patterns from an individual listening activity. For example, by filtering appropriate months and days, one could find out the approximate location where the interviewee spent his holiday season, duration of the trip, and so on. Cross-filtering also highlighted the music genres that were listened to in this peculiar time-space frame, possibly different from the person usual behavior -- and understanding the reason for this discrepancy could be addressed in the interview, e.g. listening during family meeting.
    This prototype was conceived as a proof of concept, and proved useful in encouraging our users to exchange ideas about the types of behaviors and the individual's evolution of taste that could be visualized analyzed thanks to the streaming data. It also helped the researchers in evaluating the potentials, limits and bias of the streaming history data.
    \smallskip
    \item \textbf{Winnowing} : The second step aimed at helping AMPLI's prospective users to express their visualization needs. To make them comparable, all interviews are led according to a reference "interview guide" that was conceived collaboratively. The guide consists in a series of topics (and questions within) to address with each interviewee, not necessarily in a fixed order. Interviews are conducted by six different social scientists, each of them being specifically interested in a particular topic. Those include music tastes and distastes; the biographical approach to taste formation; the use of music recommender systems; how people organize their music collections on platforms (playlists, etc.); or the different social contexts associated to different genres of music. To gather more qualitative material, social scientists collectively elaborated a "modular interview guide" that includes a common set of questions for each topic/module, and more detailed questions that will be addressed or not depending on who conducts the interview. During the making of this guide, interviewers were asked to directly suggest visualizations -- or alternatively questions that visualization could answer -- for each module. This resulted in a list of 3 to 5 visualization requests per module. The interviewers later refined these lists into actual visualization proposals.
    \smallskip
    \item \textbf{Deployment} : This consists first in developing AMPLI, and then in improving it continuously thanks to the feedback provided by its users. Social scientists first used AMPLI during a wave of thirty individual interviews that were conducted in Spring 2021. Before the wave started, each individual visualization item was reviewed to ensure that it fulfilled the requirements of the interviewers who had asked for it. Individual interviews were then prepared and conducted with AMPLI.
    We then asked our users for their feedback on how exactly they used the tool, at which moments of the interview they used it, which parts of the app revealed the most useful and which were skipped, etc.
\end{itemize}

For the implementation, we chose to rely upon a programming language and selected visualization libraries, rather than using an interactive, all-in-one software like Tableau. This is mainly because of the strong coupling between the pre-processing of the respondents' streaming data, the queries performed for each visualization, and the graphical rendering engine.
We chose to use the \texttt{R} software \cite{r_core_team_r_2021}, both for its advanced analytical and visualization features, and for its ecosystem of libraries, including the \texttt{shiny}  \cite{chang_shiny_2017} framework AMPLI is built upon. Using a versatile programming language, we can integrate all the tasks into a single "visualization pipeline"~\cite{keim_mastering_2010}, from data storage to graphic rendering on the interactive user interface\footnote{Visualizations are designed using the packages \texttt{ggplot2} \cite{wickham_ggplot2_2016} and \texttt{echarts4r} \cite{coene_echarts4r_2021}.}.

\section{An overview of AMPLI}

\begin{figure*}[!htb]
    \centering
    \includegraphics[width=.7\textwidth]{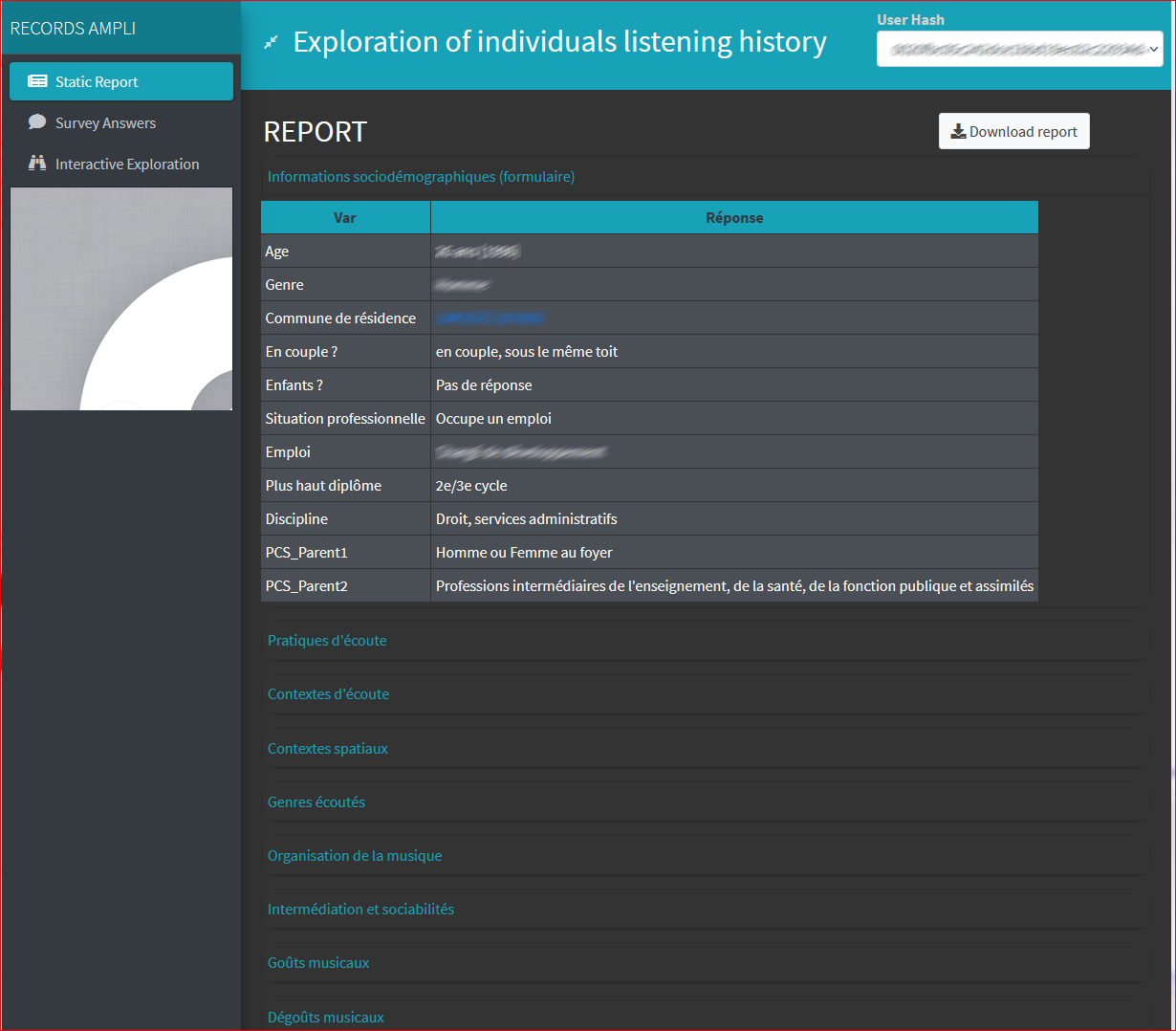}
    \caption{\label{fig:ampli_overview}The "static report" tab of AMPLI. This page provides a synthetic report on a surveyed individual. It includes both its responses to the RECORDS survey questionnaire and a summary of its listening activity on the streaming service. In this screenshot, all the "modules" of the interview guide can be seen. Using an accordion design, the numerous tables and plots of each module can be hidden, as it is shown here (only the module "Sociodemographic informations" is visible).}
    \Description{This figure displays one of the three tabs of the AMPLI application, that is the static report. It displays a table containing sociodemographic information about a survey respondent : his/her age, gender, home location, marital status and other additional information.}
\end{figure*}

We named the application AMPLI -- for "Amplifying Music listening Practices studies with Logs-augmented Interviews". It is a web application that runs on a distant server and is accessed inside the browser. It allows to explore the data of a single individual that gave his/her full written consent to participate in the RECORDS research program. More precisely, the individuals whose data can be queried through the app gave their written consent for (i) their web survey responses data to be linked with their several-year listening history data collected by the streaming platform (ii) agreed to be contacted for a recorded interview. Once logged in, the user can select the consenting respondent whose data will be explored. The application window is divided in two parts: the data visualization area in the center, and a control panel on the left from which the user can select among three available tabs (see Figure~\ref{fig:ampli_overview}). The left panel also includes a music player from which the user can play/pause the songs that can be selected in the tabs. The three tabs correspond to different use cases.

\begin{figure*}[!htp]
    \centering
    \includegraphics[width=.9\textwidth]{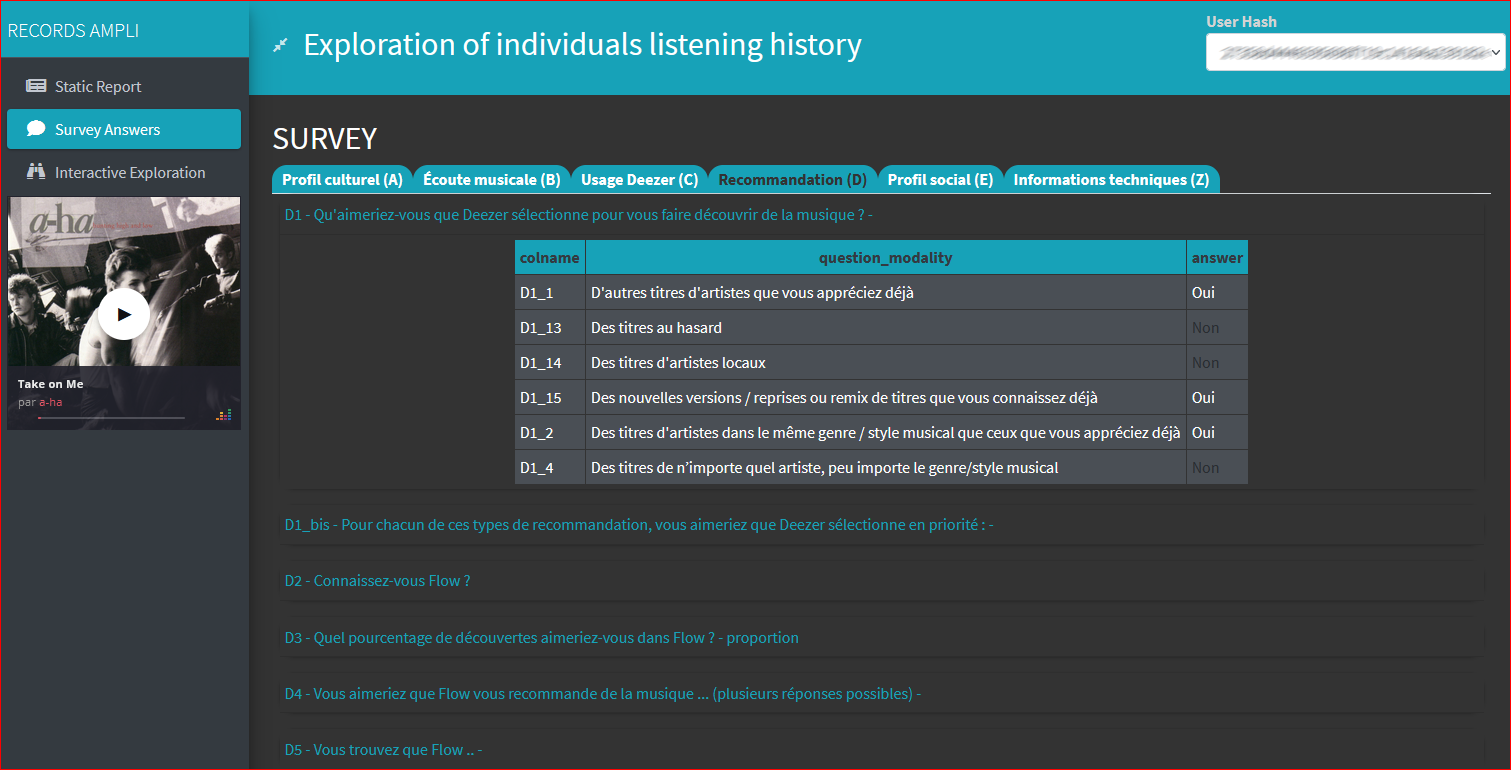}
    \caption{An example of AMPLI's survey part - The respondant's answers to this question ("How would you like Deezer to help you discover new songs ?") are highlighted to speed up the interviewee's answers analysis.}
    \label{fig:ampli_survey}
    \Description{This figure displays a screenshot of the survey part of the AMPLI application. It shows a survey respondant answers to the question "How would you like Deezer to help you discover new songs ?". Positive answers (by proposing "other tracks of already liked artists", "new versions/remixes of already known songs" and "songs with similarities of already liked ones") are highlighted in white (on a black background) while negative ones are grayed out.}
\end{figure*}

\begin{figure*}[!htp]
    \centering
    \includegraphics[width=.9\textwidth]{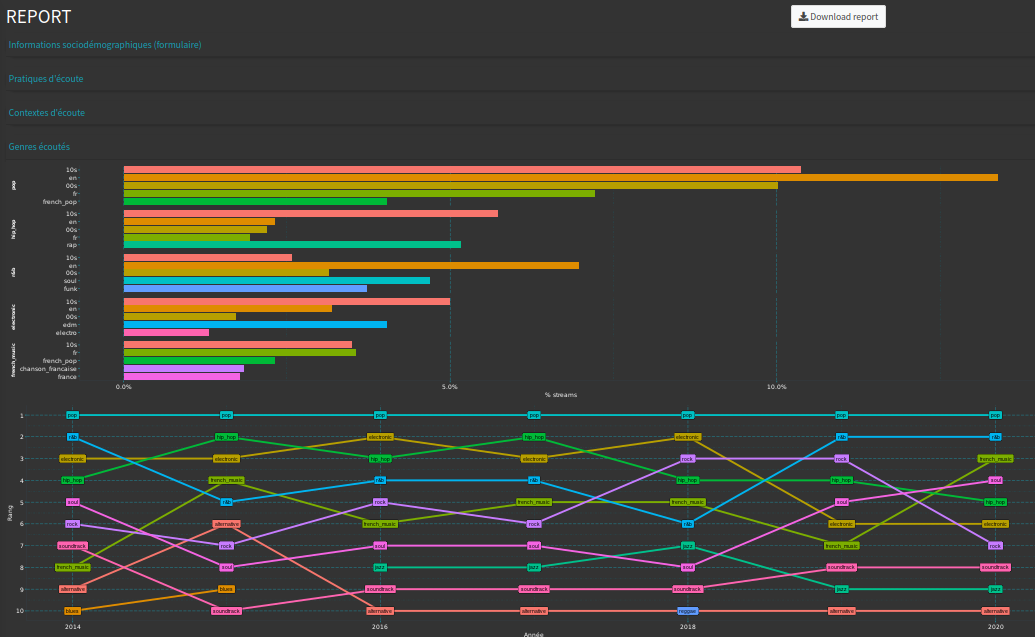}
    \caption{AMPLI survey part - We show here the time evolution of the musical genres listened to by a given interviewee on the streaming platform. It shows some volubility within the listened genres over time, the most noticeable being that while pop music is always the main and dominant genre (top plot) of the interviewee, the relative position of the other genres vary a lot over the years (middle plot), like electronic music varying between 2nd and 6th position.}
    \label{fig:ampli_genres}
    \Description{This figure displays the "Listened musical genres" module of the static report part of AMPLI. Two plots are displayed. The first one is a bar-plot of this user's most listened genres (pop, hip-hop, r&b, etc.) and sub-genres (refined by decade, music language, etc.). The second plot, commented in the caption, is a line-plot of the relative ranks of each musical genre for each year. It shows, for example, that the relative importance of electronic music in the user's listening has changed a lot over the years (the second most listened genre of the user in 2016 and 2018, but only the 6th in 2019 and 2020).
    }
\end{figure*}

\begin{itemize}
    \item \textbf{\textit{Report}} : the main tab displays a series of plots and tables that allow the user to get a general overview of the respondent's profile, thanks to his/her responses to the web survey -- including self-declared sociodemographics information -- and his/her listening behavior on the streaming platform. Content organization in this tab mirrors the structure of the interview guide, and consists in independent "modules" (e.g. see the "sociodemographics" module in Fig.~\ref{fig:ampli_overview} and the "listened genres" module in Fig.~\ref{fig:ampli_genres}). The order of appearance of the modules in the window was decided collectively. We conceived this part as a static report to be used when preparing an interview and to keep notes during it, so interactivity is limited on purpose. Following users requests, more interactivity was introduced in the "tastes" and "distastes" modules, to allow interviewers to listen to songs that are typical of the respondent's listening activity (as seen in Fig.~\ref{fig:ampli_survey}, with \textit{a}-h\textit{a} being the most listened artist of this respondent).
    This feature ensures an immediate immersion within the music of the interviewee, whose taste may be very different from the interviewer's. Indeed, being able to listen to the songs and artists the respondent played the most and the ones he/she explicitly 'like', but also the songs he/she systematically skipped or even banned provides the user a much deeper understanding of the respondent's musical "universe", especially when compared to self-declared questionnaire data.

    \item \textbf{\textit{Survey}}: This tab presents the interviewee's responses to the survey he/she previously filled in a user-friendly way. We did not include this tab in the initial design, but added it after numerous requests from the users. The main reason for it is that the web survey is long (approximately 50 questions), and several questions accept multiple combinations of possible answers. In this tab, surveys answers are organized within tables that are dynamically generated and highlighted (see fig.~\ref{fig:ampli_survey}). This lets the interviewer get a quick grasp of their interviewee's answers.
    
    \item \textbf{\textit{Exploration}} : The exploration tab is the most interactive part of AMPLI. It was included after the users expressed their regret the report tab was no longer interactive, and wished to retrieve some of the features of the prototype (see fig.~\ref{fig:proto_ampli}). This tab is then largely based on the elicitation-phase prototype. It contains visual analysis tools like multidimensional cross-selection, drill-down or interactive graphs (see Fig.~\ref{fig:ampli_exploration}) that refine the numerous possibilities of analysis of the respondent listening history data. Such interactivity helps in identifying temporal and spatial patterns within the respondents' listening habits. It can sustain more precise, contextualized questions about the contexts in which the respondent listens to music, or about his/her musical taste. For example, the dynamic graph exploration lets the interviewer finds artists that the interviewee did not listen on the platform, but that are musically very close to the ones often listened to. This can constitute a good entry to break the ice when asking the respondents about their distastes. Another use case of the exploration tab is "fact-checking", as it is easy with visual selections to isolate patterns mentioned by the respondent during the interview (for example, a respondent talking about the local music she listened when traveling abroad). By querying the respondents' data in many possible directions, visual exploration tab helps the interviewer in checking the correspondence between what the respondent said in the interview, and what he/she did according to his/her activity data.
\end{itemize}

\clearpage

\clearpage

\section{Discussion and future work}

The first thirty interviews that were conducted with AMPLI highlighted some unanticipated use cases that will guide the future developments. 
Firstly, some of the interviewers shared some selected parts of AMPLI's graphical interface with the people they interviewed, in order to elicit their feelings and reactions on topics that are known to be hard-to-grasp during traditional, unequipped sociological interviews -- including distaste in music. This spontaneous use case resembles the "personal data" visualization approach we mentioned in section~\ref{sec:review}. We are now considering the development of a lighter companion, interviewee-side app that would show selected screens from the interviewer's app. The interviewer could decide which screens he/she wants to share, and could hide some other screens to prevent "flooding" the interviewee with too much information, and limit the risks of a negative reaction to such an amount of personal data.

Another remarkable and unanticipated use-case that was reported by one of the users consisted in using AMPLI as a "fact-checker", immediately after the interview was completed, to confirm surprising assertions that were made by the respondent about his/her use of the platform.
Some AMPLI users also got back to it several months after the interviews, when transcribing and tagging the recorded audio, to retrieve some details provided by the interviewee, but that were inaudible or misunderstood, including names of songs or bands. Meanwhile, providing a general picture of an interviewee's activity and zooming on a very specific aspect of his/her activity history are two different tasks, and we are currently working on the exploration part of the application to make it suited to include both features.

These unanticipated use cases prove that a mixed methodology can greatly improve our understanding of many ordinary practices that are increasingly interfaced by digital technology. However, we are also convinced that such a methodology is not straightforward to implement, especially when the digital tracks -- user logs -- are both massive, highly multidimensional and depend on the service. How to bind together people's survey data with their digital activity data in a single interface raises many questions, some ethical and some technical. Those are fundamental questions social scientists will face in the digital age, and further interdisciplinary work and other case studies are needed for standards and good practices to emerge.

\section*{Author contributions}

RC, TL and MM designed the study. RC designed and implemented the application, and drafted the manuscript. RC, TL and MM revised and edited the manuscript. AB, JSB, SC, NL, BLB and YR conducted the interviews, used the application and provided feedback. JSB coordinated the preparation of the interviews. All authors approved the final version of the manuscript.

\begin{acks}
This work is a part of the RECORDS (\url{https://records.huma-num.fr}) research project. RECORDS is funded by the French National Research Agency (ANR) under the reference ANR-19-CE38-0013.
\end{acks}

\bibliographystyle{ACM-Reference-Format}
\bibliography{CHI2022}

\end{document}